\input harvmac
\font\cmss=cmss10 \font\cmsss=cmss10 at 7pt

\def\IB{\relax\hbox{$\inbar\kern-.3em{\rm B}$}}
\def\IC{\relax\hbox{$\inbar\kern-.3em{\rm C}$}}
\def\ID{\relax\hbox{$\inbar\kern-.3em{\rm D}$}}
\def\IE{\relax\hbox{$\inbar\kern-.3em{\rm E}$}}
\def\IF{\relax\hbox{$\inbar\kern-.3em{\rm F}$}}
\def\IG{\relax\hbox{$\inbar\kern-.3em{\rm G}$}}
\def\IGa{\relax\hbox{${\rm I}\kern-.18em\Gamma$}}
\def\IH{\relax{\rm I\kern-.18em H}}
\def\IK{\relax{\rm I\kern-.18em K}}
\def\IL{\relax{\rm I\kern-.18em L}}
\def\IP{\relax{\rm I\kern-.18em P}}
\def\IR{\relax{\rm I\kern-.18em R}}
\def\Z{\relax\ifmmode\mathchoice
{\hbox{\cmss Z\kern-.4em Z}}{\hbox{\cmss Z\kern-.4em Z}}
{\lower.9pt\hbox{\cmsss Z\kern-.4em Z}}
{\lower1.2pt\hbox{\cmsss Z\kern-.4em Z}}\else{\cmss Z\kern-.4em Z}\fi}

\def\II{\relax{\rm I\kern-.18em I}}

\def\P{{\bf P}}


\def\CN {{\cal N}}
\def\CO {{\cal O}}


\def\tilde{\widetilde}





\def\inbar{\,\vrule height1.5ex width.4pt depth0pt}



\def\pp{{\mathchoice
              %
          {
              \kern 1pt%
              \raise 1pt
              \vbox{\hrule width5pt height0.4pt depth0pt
                    \kern -2pt
                    \hbox{\kern 2.3pt
                          \vrule width0.4pt height6pt depth0pt
                          }
                    \kern -2pt
                    \hrule width5pt height0.4pt depth0pt}%
                    \kern 1pt
           }
            {
              \kern 1pt%
              \raise 1pt
              \vbox{\hrule width4.3pt height0.4pt depth0pt
                    \kern -1.8pt
                    \hbox{\kern 1.95pt
                          \vrule width0.4pt height5.4pt depth0pt
                          }
                    \kern -1.8pt
                    \hrule width4.3pt height0.4pt depth0pt}%
                    \kern 1pt
            }
            {
              \kern 0.5pt%
              \raise 1pt
              \vbox{\hrule width4.0pt height0.3pt depth0pt
                    \kern -1.9pt  
                    \hbox{\kern 1.85pt
                          \vrule width0.3pt height5.7pt depth0pt
                          }
                    \kern -1.9pt
                    \hrule width4.0pt height0.3pt depth0pt}%
                    \kern 0.5pt
            }
            {
              \kern 0.5pt%
              \raise 1pt
              \vbox{\hrule width3.6pt height0.3pt depth0pt
                    \kern -1.5pt
                    \hbox{\kern 1.65pt
                          \vrule width0.3pt height4.5pt depth0pt
                          }
                    \kern -1.5pt
                    \hrule width3.6pt height0.3pt depth0pt}%
                    \kern 0.5pt
            }
        }}
\def\mm{{\mathchoice
   %
                  %
                       {
                             \kern 1pt
               \raise 1pt    \vbox{\hrule width5pt height0.4pt depth0pt
                                  \kern 2pt
                                  \hrule width5pt height0.4pt depth0pt}
                             \kern 1pt}
                       {
                            \kern 1pt
               \raise 1pt \vbox{\hrule width4.3pt height0.4pt depth0pt
                                  \kern 1.8pt
                                  \hrule width4.3pt height0.4pt depth0pt}
                             \kern 1pt}
                       {
                            \kern 0.5pt
               \raise 1pt
                            \vbox{\hrule width4.0pt height0.3pt depth0pt
                                  \kern 1.9pt
                                  \hrule width4.0pt height0.3pt depth0pt}
                            \kern 1pt}
                       {
                           \kern 0.5pt
             \raise 1pt  \vbox{\hrule width3.6pt height0.3pt depth0pt
                                  \kern 1.5pt
                                  \hrule width3.6pt height0.3pt depth0pt}
                           \kern 0.5pt}
                       }}
\def\ad{{\kern0.5pt
                   \alpha \kern-5.05pt
\raise5.8pt\hbox{$\textstyle.$}\kern 0.5pt}}
\def\bd{{\kern0.5pt
                   \beta \kern-5.05pt \raise5.8pt\hbox{$\textstyle.$}\kern
0.5pt}}
\def\qd{{\kern0.5pt
                   q \kern-5.05pt \raise5.8pt\hbox{$\textstyle.$}\kern 0.5pt}}
\def\Dot#1{{\kern0.5pt
     {#1} \kern-5.05pt \raise5.8pt\hbox{$\textstyle.$}\kern 0.5pt}}
%


\def\comment#1{}
\def\fixit#1{}


\lref\Witten{E.~Witten, ``Non-Perturbative Superpotentials In
String Theory", Nucl.Phys. {\bf B474} (1996) 343.}
\lref\HM{J.A.~Harvey and G.~Moore,
``Superpotentials and Membrane Instantons", hep-th/9907026.}
\lref\Gukov{S.~ Gukov, ``Solitons, Superpotentials and Calibrations,''
Nucl.Phys. {\bf B574} (2000) 169.}
\lref\mirrorbook{S.-T.~Yau, editor, {\it Essays on Mirror Manifolds},
International Press, 1992; B.~Greene and S.-T.~Yau, editors,
{\it Mirror Symmetry. II}, International Press, 1997.}
\lref\OV{H.~Ooguri and C.~Vafa, ``Knot invariants and topological strings,''
Nucl. Phys. {\bf B577} (2000) 419.}
\lref\cand{P. Candelas, X.C. De la Ossa, P.S. Green and L. Parkes,
``A Pair of Calabi-Yau Manifolds As An
Exactly Soluble Superconformal Theory,'' Nucl. Phys. {\bf B359}
(1991) 21.}%
\lref\aspmor{P. Aspinwall and D. Morrison,
``Topological Field Theory and Rational Curves,''
Commun. Math. Phys. {\bf 151} (1993) 245.}%
\lref\Vafa{C.~Vafa, ``Superstrings and Topological Strings at Large N,''
hep-th/0008142.}
\lref\AV{B.~Acharya, C.~Vafa, ``On Domain Walls of N=1 Supersymmetric
Yang-Mills in Four Dimensions,'' hep-th/0103011.}
\lref\AMV{M.~Atiyah, J.~Maldacena and C.~Vafa,
``An M~theory flop as a large n duality,'' hep-th/0011256.}
\lref\atwit{M. Atiyah and E. Witten, to appear.}
\lref\CIV{F.~Cachazo, K.~Intriligator and C.~Vafa,
``A large N duality via a geometric transition,''
hep-th/0103067.}
\lref\SV{S.~Sinha and C.~Vafa, ``SO and Sp Chern-Simons at large N,''
hep-th/0012136.}
\lref\AganagicV{M.~Aganagic and C.~Vafa, ``Mirror symmetry, D-branes
and counting holomorphic discs,'' hep-th/0012041.}
\lref\AKV{M.~ Aganagic, A.~ Klemm, C.~ Vafa, ``Disk Instantons,
Mirror Symmetry and the Duality Web,'' hep-th/0105045.}
\lref\lastAV{M.~ Aganagic and C.~ Vafa, ``Mirror Symmetry and a G(2) Flop,''
hep-th/0105225.}
\lref\AcharyaSTR{B.~S.~Acharya, ``Confining strings from G(2)-holonomy
spacetimes,'' hep-th/0101206.}
\lref\AcharyaMTH{B.~S.~Acharya,``On realising N = 1 super Yang-Mills
in M~theory,'' hep-th/0011089.}
\lref\Pioline{H.~Partouche and B.~Pioline, ``Rolling among G(2) vacua,''
JHEP{\bf 0103}, 005 (2001).}
\lref\AW{M.~Atiyah and E.~Witten, to appear.}
\lref\Gomis{J.~Gomis, ``D-branes, holonomy and M~theory,''
hep-th/0103115.}
\lref\Kachru{S.~Kachru and J.~McGreevy, ``M~theory on manifolds
of G(2) holonomy and type IIA orientifolds,'' hep-th/0103223.}
\lref\Cvetic{M.~Cvetic, G.~W.~Gibbons, H.~Lu and C.~N.~Pope,
``Ricci-flat metrics, harmonic forms and brane resolutions,''
hep-th/0012011.}
\lref\CveticTRANS{M.~Cvetic, H.~Lu and C.~N.~Pope,
``Brane resolution through transgression,'' hep-th/0011023.}
\lref\CveticFRACT{M.~Cvetic, G.~W.~Gibbons, H.~Lu and C.~N.~Pope,
``Supersymmetric non-singular fractional D2-branes and NS-NS 2-branes,''
hep-th/0101096.}
\lref\CveticLTWO{M.~Cvetic, G.~W.~Gibbons, H.~Lu and C.~N.~Pope,
``Hyper-Kaehler Calabi metrics, L**2 harmonic forms, resolved
M2-branes,  and AdS(4)/CFT(3) correspondence,'' hep-th/0102185.}
\lref\CveticNEW{M. Cvetic, G.W. Gibbons, H. Lu, C.N. Pope,
``New Complete Non-compact Spin(7) Manifolds,'' hep-th/0103155.}
\lref\BS{R.L.~Bryant, and S.~Salamon, ``On the construction
of some complete metrics with exceptional holonomy,''
Duke Math. J. {\bf 58} (1989) 829.}
\lref\GPP{G.W.~Gibbons, D.N.~Page and C.N.~Pope, ``Einstein Metrics on $S^3$,
$R^3$, and $R^4$ bundles,'' Commun.Math.Phys. {\bf 127} (1990) 529.}
\lref\PT{L.A.~ Pando Zayas and A.A.~ Tseytlin, ``3-branes on resolved
conifold," hep-th/0010088.}
\lref\KKLM{S.~Kachru, S.~Katz, A.~E.~Lawrence and J.~McGreevy,
``Mirror symmetry for open strings,'' Phys.Rev.D {\bf 62} (2000) 126005;
``Open string instantons and superpotentials,'' Phys.Rev.D {\bf 62} (2000)
026001.}
\lref\TV{T.~R.~Taylor and C.~Vafa, ``RR flux on Calabi-Yau and
partial supersymmetry breaking,'' Phys.\ Lett.\  {\bf B474}, 130 (2000).}
\lref\Partouche{H.~Partouche and B.~Pioline,
``Rolling among G(2) vacua,'' JHEP {\bf 0103}, 005 (2001).}
\lref\toappear{M. Cveti\v{c}, G.W. Gibbons, H. L\"u and C.N. Pope,
``M3-branes, $G_2$ Manifolds and Pseudo-supersymmetry,'' to appear.}

\Title{\vbox{\baselineskip12pt\hbox{hep-th/0106034}
\hbox{CALT-68-2333}\hbox{CITUSC/01-022}
\hbox{ITEP-TH-27/01}\hbox{PUPT-1991}}}
{\vbox{
\centerline{Gauge Theory at Large $N$ and New $G_2$ Holonomy Metrics}
\vskip 4pt
}}
\centerline{Andreas Brandhuber,$^{\spadesuit}$ Jaume Gomis,$^{\spadesuit}$
Steven S. Gubser$^{\spadesuit \clubsuit}$ and
Sergei Gukov$^{\spadesuit \clubsuit}$}
\medskip
\vskip 8pt
\centerline{\it $^{\spadesuit}$ Department of Physics,
California Institute of Technology,}
\centerline{\it Pasadena, CA 91125}
\centerline{\it and}
\centerline{\it Caltech-USC Center for Theoretical Physics}
\centerline{\it University of Southern California}
\centerline{\it Los Angeles, CA 90089}
\medskip
\centerline{\it $^{\clubsuit}$ Joseph Henry Laboratories, Princeton University}
\centerline{\it Princeton, NJ 08544}
\medskip
\medskip
\noindent
We find a one-parameter family of new $G_2$ holonomy metrics and
demonstrate that it can be extended to a two-parameter family. These
metrics play an important role as the supergravity dual of the large $N$
limit of four dimensional supersymmetric Yang-Mills. We show that
these $G_2$ holonomy metrics describe the M~theory lift of the
supergravity solution describing a collection of D6-branes wrapping
the supersymmetric three-cycle of the deformed conifold geometry for
any value of the string coupling constant.
\smallskip
\Date{June 2001}
\newsec{Introduction}
There are at least two important reasons to study M~theory on manifolds
admitting a metric  with $G_2$ holonomy. The first one is that the
condition of four-dimensional ${\cal N}=1$ supersymmetry that follows
from the low energy approximation to M~theory -- eleven-dimensional
supergravity --  is precisely that the internal seven-dimensional
manifold admit a
$G_2$ holonomy metric. The massless four-dimensional fields that arise
from such compactifications and the classical four-dimensional
effective supergravity description can be computed from the topology of the
corresponding $G_2$ manifold
\nref\sugrakk{G.Papadopoulos and P.K. Townsend, ``Compactification of
D=11 supergravity on spaces of exceptional holonomy'',
Phys. Lett. {\bf B357} (1995) 300; J. Gutowski and G.Papadopoulos,
``Moduli Spaces and Brane Solitons for M-Theory Compactifications on
Holonomy $G_2$ Manifolds'', hep-th/0104105.}%
\sugrakk . The familiar obstruction to obtaining a chiral
four-dimensional spectrum still holds at the level of the supergravity
approximation but
non-perturbative effects -- arising for instance from singularities --
can lead to interesting models with chiral
matter and non-abelian gauge fields
from compactification of M~theory on spaces with $G_2$ holonomy.
A more recent motivation for the study of $G_2$ holonomy
manifolds is the r\^ole they play as geometric dual
descriptions of the large $N$ limit of ${\cal N}=1$ four-dimensional
gauge theories \refs{\AcharyaMTH,\AMV,\atwit}. In \AMV\  a
duality conjectured by Vafa \Vafa\
between Type IIA on the deformed conifold with D-branes and
Type IIA on the resolved conifold with Ramond-Ramond flux was derived
by lifting
the two Type IIA backgrounds to M~theory, where they take a purely
geometrical form in terms of a compactification\foot{The $G_2$ holonomy
manifolds that appear in the
M~theory lift are noncompact, so strictly speaking we are not
compactifying.} on two different
$G_2$ holonomy manifolds admitting a  smooth interpolation in
M~theory. Since the Type IIA background with D-branes naturally
contains gauge fields this duality allows one to study the infrared
dynamics of gauge theories by analyzing M~theory on
spaces with $G_2$ holonomy. This new type of duality has been
further developed and generalized in
\nref\SV{S.~Sinha and C.~Vafa, ``SO and Sp Chern-Simons at large N,''
hep-th/0012136.}%
\nref\AcharyaSTR{B.~S.~Acharya, ``Confining strings from G(2)-holonomy
spacetimes,'' hep-th/0101206.}%
\nref\AV{B.~Acharya, C.~Vafa, ``On Domain Walls of N=1 Supersymmetric
Yang-Mills in Four Dimensions,'' hep-th/0103011.}%
\nref\CIV{F.~Cachazo, K.~Intriligator and C.~Vafa,
``A large N duality via a geometric transition,''
hep-th/0103067.}%
\nref\Gomis{J.~Gomis, ``D-branes, holonomy and M~theory,''
hep-th/0103115.}%
\nref\edlnun{J.D. Edelstein and C. Nu\~nez, ``D6 branes and M~theory
geometrical transitions from gauged supergravity'', JHEP {\bf 0104}
(2001) 028.}%
\nref\Kachru{S.~Kachru and J.~McGreevy, ``M~theory on manifolds
of G(2) holonomy and type IIA orientifolds,'' hep-th/0103223.}%
\nref\Edelstein{J.~D.~Edelstein, K.~Oh and R.~Tatar,
``Orientifold, geometric transition and large N duality for
SO/Sp gauge  theories,'' JHEP {\bf 0105}, 009 (2001).}%
\nref\Kaste{P.~Kaste, A.~Kehagias and H.~Partouche,
``Phases of supersymmetric gauge theories from M~theory on G(2) manifolds,''
JHEP {\bf 0105}, 058 (2001).}%
\nref\Dasgupta{K.~Dasgupta, K.~Oh and R.~Tatar,
``Geometric transition, large N dualities and MQCD dynamics,''
hep-th/0105066. }%
\SV-\Dasgupta .

Only three examples of complete metrics with
$G_2$ holonomy are known in the literature \refs{\BS,\GPP}.
However, supersymmetry together with the
familiar Type IIA duality with M~theory indicate
\refs{\AcharyaMTH,\AMV,\Gomis,\edlnun} that there must exist a
large class of $G_2$ holonomy manifolds describing the M~theory lift of Type
IIA D6-branes wrapped on a special Lagrangian three-cycle of a Calabi-Yau
three-fold. Constructing new metrics with $G_2$ holonomy is therefore an
important enterprise which might eventually
lead to interesting four-dimensional
supersymmetric chiral models.
Moreover, new $G_2$ holonomy metrics can lead to an improved
understanding of the strongly coupled infrared dynamics of gauge
theories.
The search for new complete metrics of exceptional
holonomy was revived recently and the list of examples with
$Spin(7)$ holonomy \refs{\BS,\GPP} was extended
\nref\sup{M.Cvetic, G.W. Gibbons, H. Lu and C.N. Pope, ``Hyper-Kahler
Calabi Metrics, $L^2$ Harmonic Forms, Resolved M2-branes, and
$AdS_4/CFT_3$ Correspondence'',  hep-th/0102185.}%
\nref\CveticNEW{M. Cvetic, G.W. Gibbons, H. Lu, C.N. Pope,
``New Complete Non-compact Spin(7) Manifolds,'' hep-th/0103155.}%
\refs{\sup,\CveticNEW}. See also
\nref\Konishi{Y.~Konishi and M.~Naka,
``Coset construction of Spin(7), G(2) gravitational instantons,''
hep-th/0104208.}
\Konishi\ in which a somewhat different approach is pursued.

In this paper we construct a new metric with $G_2$ holonomy. The
metric we find describes the M~theory lift of a configuration of
D6-branes wrapping
the ${\bf S}^3$ of the deformed conifold geometry for a finite
asymptotic
value of the string coupling. The metric considered in \AMV\
describes the uplift in the limit where the string coupling is
infinite far from the D6-branes. Our
solution, as opposed to the previously known metric which is
asymptotically conical, has
at infinity a circle of finite radius which we identify with the
M~theory circle. In the interior, our new metric just reduces to
one of the previously known ones.

The plan of the rest of the paper is as follows. In section $2$ we use
symmetries and general properties of $G_2$ holonomy manifolds to write
down the metric ansatz and find a set of first order equations whose
solutions give rise to a $G_2$ holonomy metric. We analyze our system
of equations in two different interesting limits and  in this way we
recover the known $G_2$ holonomy metric  on the spin bundle over ${\bf
S}^3$ and the $SU(3)$ holonomy metric on the deformed conifold. In
section $3$ we find explicitly
a one-parameter family of new metrics with $G_2$
holonomy. We analyze its geometry and asymptotic behavior.
Furthermore, we provide convincing evidence that there exists a
two-parameter family of metrics using perturbative arguments.
In section $4$ we follow a different route based on Ricci flatness
to rederive the system of first order
equations found in section
$2$ as the conditions for $G_2$ holonomy.
Section $5$ considers the reduction of our one-parameter family
of metrics down to a Type IIA solution. By analyzing the solution we
show that it describes the supergravity background corresponding to
wrapped D6-branes on the deformed conifold.
In sections $6$ and $7$ we discuss dynamical aspects of $\CN=1$ gauge
theory on D6-branes. In particular, in section $6$ we present
a puzzle where a mismatch of the massless spectrum between Type IIA
and M~theory is exhibited.
In section $7$ we compute the non-perturbative superpotential
directly in M~theory by counting membrane instantons,
including the contributions of multiple covers.
In section $8$ we find a system of first
order equations whose solutions give rise to $G_2$ holonomy metrics
of reduced symmetry. We were not able to find metrics of this type
in closed form but by analogy with the well known Taub-NUT and
Atiyah-Hitchin metrics it is plausible that solutions to these
equations describe M~theory lifts of Type IIA orientifold six-planes
on the deformed conifold.
Finally, section $9$ contains a discussion of  our
results and further remarks on the duality studied by \AMV .
\newsec{$G_2$ Holonomy Metric and Special Solutions}
\subsec{Symmetries and Ansatz}
We want to find a metric with $G_2$ holonomy on a seven-dimensional
manifold which describes the M~theory lift of $N$ D6-branes wrapping the
${\bf S}^3$ in the deformed conifold geometry. The starting point
to accomplish
this is to write down the most general metric ansatz with a prescribed
symmetry. One way to determine the appropriate
symmetry of our ansatz is to notice that the symmetry of
the Type IIA configuration that we want to describe in eleven
dimensions is $SU(2)\times SU(2)\times Z_2$. This can be
easily understood as follows. The deformed conifold  is
described by the
following equation in ${\bf C}^4$
\eqn\coni{
z_1^2+z_2^2+z_3^2+z_4^2=r.}
This equation has an obvious $SO(4)~\sim SU(2)\times SU(2)$ symmetry
which rotates the $z_i$'s. Moreover when  $r=0$, where the space
develops a conical singularity, \coni\ has a $U(1)$  symmetry
which acts by a common phase rotation $z_i\rightarrow e^{i\alpha}
z_i$. When $r\neq 0$, then the $U(1)$ symmetry is broken to
$Z_2$, which acts by $z_i\rightarrow - z_i$. Therefore, the deformed
conifold geometry with wrapped D6-branes has an $SU(2)\times SU(2)\times Z_2$
symmetry. Furthermore, once this background is lifted to M~theory
there is an extra $U(1)$ symmetry which acts by shifts on the M~theory
circle.
Therefore, the
symmetry that we are going to impose on our ansatz for the new $G_2$
holonomy metric is going to be $SU(2)\times SU(2)\times U(1)\times
Z_2$.
Another way to understand why this is the appropriate symmetry to
impose on the metric ansatz is to notice that the geometry
found in \AMV\ describing the infinite string coupling limit of the
wrapped D6-branes on the deformed conifold has an $SU(2)\times
SU(2)\times SU(2)\times Z_2$ symmetry. If one wants a circle of finite
radius at infinity corresponding to a finite value of the string
coupling and not an asymptotically conical geometry in which
the M~theory circle decompactifies then one
of the $SU(2)$'s in the metric of \AMV\ must be broken to $U(1)$.
It is convenient to realize our ansatz  with the required symmetry
by using two sets of
left-invariant $SU(2)$ forms. In this basis the precise implementation of the
symmetry will be as $SU(2)_L\times \widetilde{SU(2)}_L\times
U(1)_R^{\hbox{diag}}\times Z_2$, where $SU(2)_L$ and
$\widetilde{SU(2)}_L$ are associated to the
two sets of left invariant one-forms
\eqn\lsigmas{\eqalign{
\sigma_1 &= \cos \psi d \theta + \sin \psi \sin \theta d \phi, \cr
\sigma_2 &= - \sin \psi d \theta + \cos \psi \sin \theta d \phi, \cr
\sigma_3 &= d \psi + \cos \theta d \phi}
\quad \quad {\rm and} \quad \quad
\eqalign{
\Sigma_1 &= \cos \tilde \psi d \tilde \theta
+ \sin \tilde \psi \sin \tilde \theta d \tilde \phi, \cr
\Sigma_2 &= - \sin \tilde \psi d \tilde \theta
+ \cos \tilde \psi \sin \tilde \theta d \tilde \phi, \cr
\Sigma_3 &= d \tilde \psi + \cos \tilde \theta d \tilde \phi }}
which satisfy the $SU(2)$ algebra
\eqn\sutwo{
d\sigma_a=-{1\over 2}\epsilon_{abc}\ \sigma_b\wedge \sigma_c\qquad
d\Sigma_a=-{1\over 2}\epsilon_{abc}\ \Sigma_b\wedge \Sigma_c.}
Then
$U(1)_R^{\hbox{diag}}=(U(1)_R\times\widetilde{U(1)}_R)_{\hbox{diag}}$
acts by a diagonal rotation on the
two sets of left-invariant $SU(2)$ forms. If we  embed the
$U(1)_R^{\hbox{diag}}$ along the Cartan generator of $SU(2)$
and $\widetilde{SU(2)}$ then  $U(1)_R^{\hbox{diag}}$ acts by a
rotation
\eqn\rotsa{
\pmatrix{
\sigma_1\cr
\sigma_2}\rightarrow \pmatrix{ \cos \alpha& -\sin \alpha \cr
\sin \alpha& \cos \alpha}\pmatrix{
\sigma_1\cr
\sigma_2}}
and likewise on $(\Sigma_1,\Sigma_2)$. Lastly, the $Z_2$
symmetry acts by exchanging the two sets of left-invariant one-forms
\eqn\ztwo{{ Z}_2 \colon~~~~ \sigma_a \leftrightarrow \Sigma_a.}
Then, the most general metric ansatz compatible with these symmetries
is given by
\eqn\metri{
ds^2 = \sum_{a=1}^7 e^a\otimes e^a,}
with the following vielbeins
\eqn\siebenbein{\eqalign{
e^1 & = A(r) (\sigma_1-\Sigma_1) ~,~~e^2 = A(r) (\sigma_2-\Sigma_2) ~, \cr
e^3 & = D(r) (\sigma_3-\Sigma_3) ~,~~e^4 = B(r) (\sigma_1+\Sigma_1) ~, \cr
e^5 & = B(r) (\sigma_2+\Sigma_2) ~,~~e^6 = C(r) (\sigma_3+\Sigma_3) ~, \cr
e^7 & = dr/C(r),}}
where we have made a particular choice for the radial coordinate. The
ansatz depends on four functions.
\subsec{$G_2$ Holonomy and First Order Equations}
Local reduction of the holonomy group of a seven-manifold from $SO(7)$
to $G_2$ is determined by the $G_2$-structure. This is a globally
defined three-form
$\Phi$, usually called the associative three-form, which is
covariantly constant with respect to the Levi-Civita connection
determined by the metric.  An equivalent statement is that it is closed and
co-closed:\foot{This is actually a stronger condition than that
$\Phi$ be harmonic, because on a non-compact manifold, closed and
co-closed implies harmonic, but not the converse.}
\eqn\harm{\eqalign{
d\Phi&=0\cr
d*\Phi&=0 \,.}}
The choice of a $G_2$-structure on the seven-dimensional manifold
breaks the  $GL(7,\IR)$ tangent space symmetry
to precisely $G_2$. This follows because locally, given a vielbein basis $e^a$,
where $a=1,\ldots,7$, the associative three-form is given by
\eqn\assocform{
\Phi = {1 \over 3!}\psi_{abc}\ e^a e^b e^c}
where $\psi_{abc}$ -- which are totally antisymmetric --
are the structure constants of the imaginary
octonions,
\eqn\octo{i_a i_b = -\delta_{ab} + \psi_{abc}\ i_c ~,~~ a,b,c=1, \ldots 7~}
and $G_2$ is the automorphism group of the imaginary
octonions. In a choice of basis the non-zero structure constants are
given by
\eqn\octoconst{ \psi_{abc} = +1 ~,~~ (abc) = \{(123),
(147), (165), (246), (257), (354), (367) \} ~. }
Likewise, the coassociative four-form $*\Phi$ is locally given by
\eqn\coassform{ *\Phi = {1 \over 4!}\psi_{abcd}\ e^a e^b e^c e^d }
where the totally antisymmetric structure constants $\psi_{abcd}$ in
the basis \octoconst\ are given by
\eqn\psifour{\psi_{abcd} = +1 , (abcd) = \{(4567), (2356), (2374),
(1357), (1346), (1276), (1245)\} }
and also have $G_2$ symmetry.
We now use the vielbeins \siebenbein\ of our ansatz to construct by
means of \assocform\ and \coassform\
$\Phi$ and $*\Phi$. This construction gives a candidate
$G_2$-structure $\Phi$. To prove that our metric has $G_2$ holonomy we
must impose that the associative three-form $\Phi$ is closed and co-closed.
These conditions imposed on the
associative three-form $\Phi$ constructed from our vielbeins
leads to the following system of first
order differential equations\foot{For completeness, we have included
in the Appendix the
exterior calculus of the vielbeins of the ansatz \siebenbein\ needed
to verify the calculations in this paper.}
\smallskip
\eqn\BPSeqns{\eqalign{
{d A \over dr} &= {1 \over 4} \left[ {B^2-A^2+D^2 \over B C D} +
   {1 \over A} \right] \cr
{d B \over dr} &= {1 \over 4} \left[ {A^2-B^2+D^2 \over A C D} -
   {1 \over B} \right] \cr
{d C \over dr} &= {1 \over 4} \left[ {C \over B^2} - {C \over A^2} \right] \cr
{d D \over dr} &= {1 \over 2} \left[ {A^2+B^2-D^2 \over A B C} \right]
.}}
\smallskip
\noindent
The general solution to these equations gives rise to a metric with
$G_2$ holonomy. In section $3$ we find a solution of these equations
in terms of elementary functions.
We conclude this subsection with two clarifying remarks. First, the
conditions \harm\ actually only
guarantees that the holonomy group of the seven manifold is contained
in $G_2$. Therefore,
we need a further criterion which determines when the holonomy group is
precisely
$G_2$. Such a criterion is known \BS, and it requires that
there exist no non-zero covariantly constant one-forms on the seven manifold
or equivalently that the fundamental group of the manifold be pure
torsion. Informally, this condition is the statement that the
seven-manifold cannot be written as the direct product of two
spaces. As will be clear
in the next section when we analyze the geometrical
properties of our ansatz the holonomy of our
metric is precisely $G_2$.
Another important point to keep in mind is as follows. As explained
above, the choice of a $G_2$-structure on the manifold breaks the
tangent space symmetry to $G_2$. On the other hand, the choice of a
metric $g$ on the manifold breaks the tangent space symmetry to
$SO(7)$. Therefore, a given $G_2$-structure $\Phi$ uniquely
determines a $G_2$ holonomy metric $g$ since $G_2\in SO(7)$.  The
converse is not true,
given a metric $g$ there is no canonical choice of a $G_2$-structure. In
the case under study, we have succeeded in constructing the associative
three-form from the metric because we have used a very suitable
choice of vielbeins for the metric. Generically it is difficult to
choose the right set of vielbeins of the metric to construct the
associative three-form. In section $4$ we follow a more traditional
approach where
we rederive the first order differential equations \BPSeqns\ from the
second order equations derived by imposing Ricci flatness on the
metric ansatz \metri\siebenbein . In this
approach one has to prove that the metric has $G_2$ holonomy either by
constructing the $G_2$-structure or by showing that there is only one
component of the $SO(7)$ spinor which is
covariantly constant.\foot{The covariantly constant spinor is
the singlet in the decomposition of the spinor
of $SO(7)$
${\bf 8}\rightarrow {\bf 1}\oplus {\bf 7}$ under $G_2$.} We have presented
the non-canonical approach first because it is the fastest way to
getting the first order equations for the metric and because by
construction it guarantees that the metric has holonomy contained in
$G_2$.
\subsec{Special Solutions}
In this section we specialize our ansatz \metri\siebenbein\
to have an enhanced $SU(2)\times
SU(2)\times SU(2)\times Z_2$  symmetry and we show that the
metric we get is the previously known $G_2$ holonomy metric on the
spin bundle of
${\bf S}^3$. We also analyze our system of first order
equations when one of the functions vanishes. In this way we recover
the known metric of $SU(3)$ holonomy on the deformed conifold geometry
\nref\ossa{P.~Candelas and X.~C.~de la Ossa, ``Comments On Conifolds,''
Nucl.\ Phys.\ B {\bf 342}, 246 (1990).}
\ossa.
By setting $A=D$ and $B=C$ in our metric ansatz \metri\siebenbein\ the
metric acquires an  enhanced $SU(2)\times
SU(2)\times SU(2)\times Z_2$  symmetry. Then the metric takes the
following form
\eqn\extrasymm{
ds^2=A^2\sum_{a=1}^3(\sigma_a-\Sigma_a)^2+
B^2\sum_{a=1}^3(\sigma_a+\Sigma_a)^2+dr^2/B^2.}
In order to understand the geometry it is useful to
make the following coordinate transformation.\foot{This transformation
has been independently studied in a forthcoming paper \toappear.}
Let $U$ and $V$ be
the $SU(2)$ group elements from which one constructs the left invariant
$SU(2)$ one-forms $\sigma_a$ and $\Sigma_a$
\eqn\construct{\eqalign{
\sigma&=\sigma_a T^a =U^{-1}dU\cr
\Sigma&=\Sigma_a T^a =V^{-1}dV,}}
where $T^a$ are the $SU(2)$ group generators. Then one can construct
the following new set of left-invariant $SU(2)$ forms
\eqn\newset{
\widetilde{w}=V(\sigma-\Sigma)V^{-1}.}
where $\widetilde{w}=\widetilde{W}^{-1}d{\widetilde W}$ for
${\widetilde W}=U
V^{-1}$. Moreover it follows
from \newset\ that the following relation holds
\eqn\plurela{
V(\sigma+\Sigma)V^{-1}={\widetilde w}-2w,}
where $w$ is a left-invariant $SU(2)$ one-form such that $w=W^{-1}dW$
with $W=V^{-1}$. With this new set of left invariant one forms $w$ and
$\widetilde{w}$ the metric \extrasymm\ takes the simple form
\eqn\recognize{
ds^2=A^2\sum_{a=1}^3\widetilde{w}_a^2+
4B^2\sum_{a=1}^3(w_a-{1\over 2}\widetilde{w}_a)^2+dr^2/B^2.}
For this more symmetric ansatz, when $A=D$ and $B=C$, our first order
equations \BPSeqns\
reduce to the
following simple ones
\eqn\simpli{\eqalign{
{dA\over dr}&={1\over 2A}\cr
{dB\over dr}&={1\over 4B}\left(1-{B^2\over A^2}\right)}}
which agree with the first order equations for the metric
on the spin bundle over
${\bf S}^3$ \refs{\GPP,\sup}.
Solution to these equations yields the following $G_2$ holonomy metric
\refs{\BS,\GPP}
\eqn\olgtwomet{
ds^2={\rho^2\over 12}\sum_{a=1}^3\widetilde{w}_a^2+
{\rho^2\over 9}(1-{\rho^3_0\over
\rho^3})\sum_{a=1}^3(w_a-{1\over 2}\widetilde{w}_a)^2+{d\rho^2/
(1-{\rho^3_0\over
\rho^3})}}
where $\rho\geq\rho_0$ and $\rho_0$ determines the size of the ${\bf
S}^3$ generated by the ${\widetilde w_a}'s$ at $\rho=\rho_0$. This
space is topologically ${\bf R}^4\times{\bf S}^3$ so the space is
simply connected which together with the explicit construction of the
$G_2$-structure \assocform , which is given by
\eqn\assomax{
\sqrt{3} \Phi={\rho_0^3\over
144}\epsilon_{abc}\;\widetilde{w}_a\wedge\widetilde{w}_b\wedge\widetilde{w}_c+
{\rho^2\over 6}d\rho\wedge \widetilde{w}_a\wedge w_a+
{(\rho^3-\rho_0^3)\over
36}\epsilon_{abc}\;(\widetilde{w}_a\wedge\widetilde{w}_b\wedge w_c -
\widetilde{w}_a\wedge w_b\wedge w_c)}
guarantees that \olgtwomet\ has $G_2$ holonomy.
We note
for future reference that this metric is asymptotically conical and
the base of the cone is topologically ${\bf S}^3\times{\bf S}^3$.

It is also interesting to consider our ansatz \metri\siebenbein\
when $C=0$. Then the system of equations \BPSeqns\ reduces to
\eqn\BPSeqnsconi{\eqalign{
{d A \over dt} &= {1 \over 4} \left[ {B^2-A^2+D^2 \over B  D}\right] \cr
{d B \over dt} &= {1 \over 4} \left[ {A^2-B^2+D^2 \over A  D}\right] \cr
{d D \over dt} &= {1 \over 2} \left[ {A^2+B^2-D^2 \over A B } \right]
,}}
where the new radial coordinate $t$ is related to the old one by $dr=C
dt$ and the metric of the resulting six-dimensional manifold looks like
  \eqn\metcon{
   ds^2=A^2\left((\sigma_1-\Sigma_1)^2+(\sigma_2-\Sigma_2)^2\right)+
     B^2\left((\sigma_1+\Sigma_1)^2+(\sigma_2+\Sigma_2)^2\right)+
     D^2(\sigma_3-\Sigma_3)^2+dt^2 \,.
  }
 We should note that setting $C=0$ reduces the symmetry of the ansatz
to just $SU(2)\times SU(2)\times Z_2$ which is precisely the symmetry
of the deformed conifold geometry \ossa.
In order to identify the geometry we rewrite the metric in a
way that is conducive to comparison with the known deformed
conifold metric. It is straightforward to change coordinates to rewrite
the metric \metcon\
in the following way:
\eqn\defconm{
ds^2={B^2 \over 2} ((g^1)^2+(g^2)^2)+
  {A^2 \over 2}((g^3)^2+(g^4)^2)+D^2(g^5)^2+dt^2,}
where
\eqn\basis{
g^1=E^1-E^3,\quad g^2=E^2-E^4,\quad g^3=E^1+E^3,\quad
g^4=E^2+E^4,\quad g^5=E^5}
with
\eqn\defco{\eqalign{
E^1&=-\sin\theta_1d\phi_1\cr
E^2&=d\theta_1\cr
E^3&=\cos\psi_1\sin\theta_2d\phi_2-\sin\psi_1 d\theta_2\cr
E^4&=\sin\psi_1\sin\theta_2 d\phi_2+\cos\psi_1 d\theta_2\cr
E^5&=d\psi_1+\cos\theta_1 d\phi_1+\cos\theta_2 d\phi_2 \,.}}
This is a popular set of vielbeins in which to write the deformed conifold
metric. The metric as written in \defconm\ is the familiar ansatz
for the deformed conifold metric.
In order to show that the $C=0$ truncation of our first order
equations indeed describes the deformed conifold we must show that the
left-over functions $A,B$ and $D$ equations \BPSeqnsconi\ coincide
with the first order equations of
the deformed conifold. A straightforward calculation shows that the
$C=0$ truncation of our metric indeed yields the deformed conifold
first order equations
\nref\pantsy{L.A. Pando Zayas and A.A. Tseytlin, ``3-branes on
resolved conifold'', hep-th/0010088.}%
\pantsy\ whose solution gives the $SU(3)$ holonomy
metric \ossa\ which in a
convenient choice of the radial coordinate is given by
\eqn\finmet{\eqalign{
ds^2=K(\tau)\Big[{1\over
3K^3(\tau)}(d\tau^2+(g^5)^2) & +{1\over 4}{\sinh^2\left({\tau\over
2}\right)}[(g^1)^2+(g^2)^2] + \cr
& +{1\over 4}{\cosh^2\left({\tau\over 2}\right)}[(g^3)^2+(g^4)^2]\Big]}}
with
\eqn\kfunc{
K(\tau)={(\sinh(2\tau)/2-\tau)^{1/3}\over \sinh(\tau)}.}
Asymptotically this metric is also conical and the base of the cone is
topologically ${\bf S}^2\times {\bf S}^3$.
This is quite satisfactory since setting $C=0$, roughly
speaking, removes the twisting on the M~theory circle due to the
wrapped D6-branes on the ${\bf S}^3$ of the deformed conifold. Once
the twisting is removed one just gets the unperturbed deformed conifold
metric as we have just shown. When $C\neq 0$ one gets a non-trivial
$U(1)$ fibration which we analyze next.
\newsec{A New Complete $G_2$ Holonomy Metric and Its Geometry}
As shown in section $2$ the most general metric on a seven-dimensional
manifold
having  $SU(2)\times SU(2)\times U(1)\times Z_2$ symmetry
depends on four
functions and the requirement of having a $G_2$-structure resulted in the
following system of coupled first order equations
\smallskip
\eqn\BPSeqnsa{\eqalign{
{d A \over dr} &= {1 \over 4} \left[ {B^2-A^2+D^2 \over B C D} +
   {1 \over A} \right] \cr
{d B \over dr} &= {1 \over 4} \left[ {A^2-B^2+D^2 \over A C D} -
   {1 \over B} \right] \cr
{d C \over dr} &= {1 \over 4} \left[ {C \over B^2} - {C \over A^2} \right] \cr
{d D \over dr} &= {1 \over 2} \left[ {A^2+B^2-D^2 \over A B C} \right]
.}}
This complicated system of
equations has the following discrete ${\bf Z}_2$ symmetry
\eqn\newztwo{
{\bf Z}_2 \colon~~~~ \cases{
r \to - r \cr
A \leftrightarrow B \cr
D \to -D.}
}
Therefore, the most general solution to \BPSeqnsa\ invariant under
this symmetry depends on three parameters.
However, one of them is trivial since it just
corresponds to shifting by a constant the radial
coordinate. Therefore, the general solution to \BPSeqnsa\ depends on
two nontrivial parameters. As we shall see later, the two parameters
can be  interpreted from the Type IIA perspective as determining the
string coupling $g_s$ and the size of the ${\bf S}^3$ in the deformed
conifold geometry on which the D6-branes are wrapped. In the eleven-dimensional
description these two parameters correspond respectively to the size
of the $U(1)$ fiber at infinity and to the volume of an ${\bf S}^3$ inside
the seven-dimensional manifold of $G_2$ holonomy.
\subsec{A family of solutions}
We have been able to find the following solution to the equations \BPSeqnsa
\eqn\coeffs{\eqalign{
A & = {1 \over \sqrt{12}} \sqrt{(r - 3 /2)(r + 9 /2)} \cr
B & = {1 \over \sqrt{12}} \sqrt{(r + 3 /2)(r - 9 /2)} \cr
C & = \sqrt{(r - 9 /2)(r + 9 /2) \over (r - 3 /2)(r + 3 /2)} \cr
D & = r/3.}}
One can then use \coeffs\ to find a one-parameter family of metrics
\metri\siebenbein\ with $G_2$ holonomy. Since \coeffs\ solve the first
order equations \harm\ which follow from the existence of a
$G_2$-structure and if we let $g$ be  the corresponding metric then
the rescaled
metric $r_0^2\; g$ also admits a $G_2$-structure. After rescaling the
radial coordinate $r\rightarrow r/r_0$ one gets the following
one-parameter family of metrics
\eqn\metricres{
ds^2 = \sum_{a=1}^7 e^a\otimes e^a,}
with the following vielbeins
\eqn\siebenbeinres{\eqalign{
e^1 & = A(r) (\sigma_1-\Sigma_1) ~,~~e^2 = A(r) (\sigma_2-\Sigma_2) ~, \cr
e^3 & = D(r) (\sigma_3-\Sigma_3) ~,~~e^4 = B(r) (\sigma_1+\Sigma_1) ~, \cr
e^5 & = B(r) (\sigma_2+\Sigma_2) ~,~~e^6 = r_0\; C(r)
(\sigma_3+\Sigma_3) ~, \cr
e^7 & = dr/C(r)}}
where now
\eqn\coeffs{\eqalign{
A & = {1 \over \sqrt{12}} \sqrt{(r - 3 r_0/2)(r + 9 r_0/2)} \cr
B & = {1 \over \sqrt{12}} \sqrt{(r + 3 r_0/2)(r - 9 r_0/2)} \cr
C & = \sqrt{(r - 9 r_0/2)(r + 9 r_0/2) \over (r - 3 r_0/2)(r + 3 r_0/2)} \cr
D & = r/3.}}
The resulting metric is Ricci flat and complete for either $r \geq 9
r_0/2$ or $r \leq -9 r_0/2$. These two solutions are related to each
other by the action of the non-trivial ${\bf Z}_2$
automorphism \newztwo\ of the first order differential equations
\BPSeqnsa . For concreteness, we will consider from now on the
solution whose radial coordinate is constrained to be $r \geq 9
r_0/2$.
The metric is Ricci flat and has a $G_2$-structure that we construct
in terms of the vielbeins \siebenbeinres\ together
with \coeffs . The $G_2$-structure can be conveniently written as
\eqn\Gttwoexact{\eqalign{
\Phi={9r_0^3\over 16}\epsilon_{abc}\;(\sigma_a\wedge\sigma_b\wedge\sigma_c-
 \Sigma_a\wedge\Sigma_b\wedge\Sigma_c)&+d\Big({r\over
 18}(r^2-{27r_0^2\over 4})(\sigma_1\wedge
 \Sigma_1+ \sigma_2\wedge
 \Sigma_2)+\cr
&+ {r_0\over
 3}(r^2-{81r_0^2\over 8})\sigma_3\wedge
 \Sigma_3\Big).}}
The existence of this covariantly constant three-form guarantees that
our metric has holonomy contained in $G_2$.
\subsec{The Geometry of the Solution and Asymptotics}
The metric we found in the previous subsection is a $U(1)$ bundle over
a six-dimensional manifold. The circle, which is
parameterized by the vielbein $e^6$, has its size at infinity set by
$r_0$ since $C\rightarrow 1$ as $r\rightarrow \infty$. In the
interior, when $r\rightarrow 9r_0/2$, then $C\rightarrow 0$ so that
the circle shrinks to zero size. This behavior is very similar to
that of the Taub-NUT metric which is not surprising since our solution
describes the M~theory lift of a wrapped D6-brane. In particular, the
size of the circle at infinity -- given by $r_0$ --  determines the
Type IIA string coupling constant.
One can generalize the coordinate transformation presented in section
2.3 and rewrite the metric as follows
\eqn\metrcol{
ds^2=A^2((g^1)^2+(g^2)^2)+B^2((g^3)^2+(g^4)^2)+D^2(g^5)^2+r_0\ C^2 (g^6)^2
+dr^2/C^2,}
where $g^1,\ldots,g^5$ are defined in \basis\defco\ and
\eqn\gsixan{
g^6=d\psi_2+\cos \theta_1 d\phi_1-\cos\theta_2 d\phi_2.}
Then the asymptotic behavior of the metric at infinity is given by
\eqn\inftmet{
ds^2=dr^2+r^2\left({1\over 9}\left(d\psi_1+\sum_{i=1}^{2}\cos \theta_i
d\phi_i\right)^2+{1\over 6}\sum_{i=1}^2\left(d\theta_i^2+\sin^2 \theta_i
d\phi_i^2\right)\right)+r_0\; (g^6)^2.}
This geometry is that of a $U(1)$ bundle over the singular conifold
metric with $SU(3)$ holonomy. The base of the cone is described by the
Einstein metric on the homogeneous space $T^{1,1}=(SU(2)\times
SU(2))/U(1)$ where the $U(1)$ is diagonally embedded along the Cartan
generator of the $SU(2)$'s. Therefore, at infinity our metric is
topologically ${\bf R}_+\times {\bf S}^1\times {\bf S}^2\times {\bf S}^3$.

We now analyze the geometry in the interior. The metric is
non-singular everywhere and near $r = 9r_0/2$ it behaves
like
\eqn\interi{
ds^2\sim d\rho^2+{9\over 4}r_0^2\left((g^1)^2+(g^2)^2+(g^5)^2\right)+
{\rho^2\over 16}\left((g^3)^2+(g^4)^2+(g^6)^2\right),}
where $\rho^2=8r_0(r-9r_0/2)$. Therefore, there is an ${\bf S}^3$ of finite
size and topologically the space becomes ${\bf R}^4 \times {\bf S}^3$.
Since $A=D$ and $C=D$ as $r\rightarrow 9r_0/2$, in the interior our
solution has enhanced $SU(2)\times SU(2)\times SU(2)\times Z_2$ symmetry.
This is exactly the situation discussed in detail in section $2.3$.
Hence, in the interior our new metric approaches the behavior of
the previously known asymptotically conical metric on the spin bundle
over ${\bf S}^3$. In fact, both $G_2$ manifolds have the same Betti numbers.

The space on which we put the $SU(2)\times SU(2)\times U(1)\times Z_2$
metric \metricres\ -- \siebenbeinres\ is homotopic to ${\bf R}^4\times {\bf
S}^3$, so it is simply connected.  As explained in section 2.2 this
implies that the holonomy of our metric is exactly $G_2$. Therefore, the
$G_2$-structure \Gttwoexact\ provides a local reduction of the
holonomy group from $SO(7)$ to precisely $G_2$ and guarantees the
existence of a unique covariantly constant spinor.

\subsec{Evidence for a two-parameter family of solutions}

A peculiarity of our solution \coeffs\ is that the size of the
topologically non-trivial ${\bf S}^3$ in the interior and the size of
the circle at infinity are in fixed ratio: both are determined by
$r_0$. In section $5$, when we reduce our solution to Type IIA this
translates into the statement that the string coupling constant and
the size of the ${\bf S}^3$ are not independent.  On the Type IIA side,
one should be free to adjust the dilaton at infinity independently of
the size of the ${\bf S}^3$.  Thus we expect that our one-parameter
family of solutions can be generalized to a two-parameter family.  We
have not been able to find two-parameter solutions in closed form, but
numerical integration of the BPS equations \BPSeqnsa\ indicates that
they exist.  Rather than mapping out the parameter space with
extensive numerics, we will be satisfied to give here a perturbative
argument which uses numerics only to check one important point.

Suppose we start with the solution \coeffs\ and wish to make a
uniformly small perturbation.  We write
  \eqn\PerturbABCD{
   A = A_0 + \phi_A \qquad
   B = B_0 + \phi_B \qquad
   C = C_0 + \phi_C \qquad
   D = D_0 + \phi_D \,,
  }
where $A_0$, $B_0$, $C_0$, and $D_0$ are the solutions given
explicitly in \coeffs.  Plugging these expressions
into \BPSeqnsa\ and expanding to
linear order in the $\phi_i$'s, we obtain an equation of the form
  \eqn\linearPhi{
   {d\vec\phi \over dr} = {\bf M}(r) \vec{\phi}
  }
 where ${\bf M}(r)$ is some $4\times 4$ matrix whose entries can be
simply expressed in terms of $A_0$, $B_0$, $C_0$, and $D_0$.  The
asymptotic forms of ${\bf M}(r)$ near $r = 9/2$ (the rounded tip of the
cone) and $r = \infty$ will be helpful:
  \eqn\AsymptoticMOne{
   {\bf M}(r) = {{\bf m}_0 \over r-9/2} +
    {{\bf m}_1 \over \sqrt{r-9/2}} + O(1)
  }
 for $r$ close to $9/2$, where
  \eqn\LittleMs{
   {\bf m}_0 = \pmatrix{ -1 & 0 & 0 & 1  \cr
                           0 & 1/2 & -1 & 0  \cr
                           0 & -1 & 1/2 & 0  \cr
                           2 & 0 & 0 & -2 } \qquad
   {\bf m}_1 = \pmatrix{ 0 & {1 \over 3\sqrt{2}} &
                            -{1 \over 3\sqrt{2}} & 0  \cr\noalign{\vskip1\jot}
                          0 & 0 & 0 & 0  \cr
                          0 & 0 & 0 & 0  \cr\noalign{\vskip1\jot}
                          0 & {\sqrt{2} \over 3} &
                            -{\sqrt{2} \over 3} & 0 }
  }
 and
  \eqn\AsymptoticMTwo{
   {\bf M}(r) = {\bf M}_0 + {{\bf M}_1 \over r} +
    O\left( {1 \over r^2} \right)
  }
 for large $r$, where
  \eqn\BigMs{
   {\bf M}_0 =
    \pmatrix{ 0 & 0 & -{1 \over 2\sqrt{3}} & 0  \cr\noalign{\vskip1\jot}
              0 & 0 & -{1 \over 2\sqrt{3}} & 0  \cr\noalign{\vskip1\jot}
              0 & 0 & 0 & 0  \cr\noalign{\vskip1\jot}
              0 & 0 & -{1 \over 3} & 0 } \qquad
   {\bf M}_1 =
    \pmatrix{ -{3 \over 2} & {1 \over 2} & {\sqrt{3} \over 2} &
                   {\sqrt{3} \over 2}  \cr\noalign{\vskip1\jot}
              {1 \over 2} & -{3 \over 2} & -{\sqrt{3} \over 2} &
                   {\sqrt{3} \over 2}  \cr\noalign{\vskip1\jot}
              0 & 0 & 0 & 0  \cr\noalign{\vskip1\jot}
              {4 \over \sqrt{3}} & {4 \over \sqrt{3}} & 0 & -4 }
  }
 Because of the $1/(r-9/2)$ term in \AsymptoticMOne, a perturbation
$\phi(r)$ will blow up logarithmically near $r = 9/2$ unless $\lim_{r
\to 9/2} \phi(r)$ is annihilated by ${\bf m}_0$.  In fact, ${\bf m}_0$
has only one null eigenvector, namely $(1,0,0,1)$, which corresponds to
expanding the unshrunk ${\bf S}^3$.  So there is only one perturbation
regular at the origin.  Because the third row of both ${\bf M}_0$ and
${\bf M}_1$ vanish completely, and the additional contributions are
$O(1/r^2)$, we see that any perturbation will lead to a finite change
in the radius of the circle at infinity.

We have now shown that the perturbation is a well-defined deformation
of the solution \coeffs.  The only remaining detail is to ensure that
it is not the deformation that we have already studied---that is, a
rigid rescaling.  A completely analytic way to check this would be to
solve \linearPhi\ in a series around $r=9/2$ and check that
$\phi_A/A_0 \neq \phi_B/B_0$ at some order in $r-9/2$.  However, it is
perhaps more to the point to demonstrate that the circle at infinity
changes its radius by a different amount from the unshrunk ${\bf
S}^3$.  It is straightforward to obtain
  \eqn\PhiRatios{
   \lim_{r \to 9/2} {\phi_A \over A_0} =
   \lim_{r \to 9/2} {\phi_D \over D_0} = {2 \over 3} \lambda \ \neq\
   \lim_{r \to \infty} {\phi_C \over C_0} \approx 0.11 \lambda \,,
  }
 for the unique perturbation regular at $r=9/2$.  Here
$\lambda$ is a parameter measuring the strength of the
perturbation, and to get the crucial $0.11$ we numerically integrated
\linearPhi\ with initial conditions $\phi_A = \phi_D = 1$, $\phi_B =
\phi_C = 0$ imposed very close to $r = 9/2$.  Thus this perturbation
does correspond to changing the unshrunk ${\bf S}^3$ by a different
scale factor from the circle at infinity---this is to be compared with
our explicitly known one-parameter family, all of which were obtained by
rigidly scaling the solution \coeffs.  Standard implicit function
arguments suffice to show that the perturbative result implies the
existence of a two-parameter family of non-singular solutions in some
finite neighborhood of the one-parameter slice which we have exhibited
in closed form.

The perturbation problem was particularly benign in this case because
$\phi_i$ could be taken uniformly small as compared to the unperturbed
solution.  By way of comparison, let us investigate another corner of
the parameter space: if the circle at infinity is very small but the
unshrunk ${\bf S}^3$ is finite, then we should be able to probe a
little ways into the full two-parameter family of solutions by
perturbing around the deformed conifold solution, \finmet.\foot{We
thank E.~Witten for a discussion which motivated us to carry out this
analysis.} Let us use a radial variable $\tau$ such that $dr = C\nu
d\tau$, where $\nu$ is some function of radius to be set at our
convenience.  The first order equations \BPSeqnsa\ become
  \eqn\BPSAgain{\eqalign{
   {dA \over d\tau} &= {\nu \over 4} \left[
    {B^2-A^2+D^2 \over BD} + {C \over A} \right]  \cr
   {dB \over d\tau} &= {\nu \over 4} \left[
    {A^2-B^2+D^2 \over AD} - {C \over B} \right]  \cr
   {dC \over d\tau} &= {\nu \over 4} \left[
    {C^2 \over B^2} - {C^2 \over A^2} \right]  \cr
   {dD \over d\tau} &= {\nu \over 2} \left[
    {A^2 + B^2 - D^2 \over AB} \right] \,,
  }}
 and we make the gauge choice $\nu = 1/\sqrt{3} K(\tau)$ so that the
unperturbed solution can be written as
  \eqn\Unperturbed{
   A = \sqrt{K \over 2} \cosh {\tau \over 2} \qquad
   B = \sqrt{K \over 2} \sinh {\tau \over 2} \qquad
   C = 0 \qquad
   D = {1 \over \sqrt{3} K} \,.
  }
 See \kfunc\ for a definition of $K(\tau)$.  We might expect to be
able to do a straightforward perturbation expansion in small $C$.
This almost works: plugging the unperturbed solution \Unperturbed\
into the third equation in \BPSAgain, one obtains
  \eqn\CEqn{
   {d \over d\tau} \left( {1 \over C} \right) =
    -{2 \over \sqrt{3}} \left( \sinh(2\tau)/2 - \tau \right)^{-2/3} \,.
  }
 While there is no elementary expression for $C(\tau)$, it is clear
that the limit $C_\infty = \lim_{\tau\to\infty} C(\tau)$ is finite and
represents the one integration constant of \CEqn.  Furthermore,
because the right hand side of \CEqn\ is everywhere negative,
$1/C(\tau)$ is monotonically decreasing in $\tau$.  Assuming $C_\infty
> 0$ we learn that $C(\tau)$ is uniformly bounded by $C_\infty$.

Solving \CEqn\ in a series around $\tau = 0$ results in
  \eqn\CSeries{
   C = {\tau \over \root 6 \of {12}} \left( 1 + O(\tau^2) \right) \,.
  }
 The problem is that near $\tau = 0$, the term $-C/B$ in the {\it
second} equation of \BPSAgain\ is no longer small compared to the
other term in square brackets.  This ``back-reaction'' of the $U(1)$
fiber on the deformed conifold geometry invalidates the
straightforward perturbation expansion in a small neighborhood of
$\tau = 0$.  This is analogous to the phenomenon of boundary layers.
One can see that there is a problem by observing that if we use
\CSeries, then $\lim_{\tau\to
0} C/B = 2$, which means that the shrinking ${\bf S}^3$ is not round.
This would actually mean that there is a singularity at $\tau = 0$.
The resolution is to give a different perturbation analysis near $\tau
= 0$ by assuming $A \sim O(1)$, $B \sim O(\tau)$, $C \sim O(\tau)$,
and $D \sim O(1)$.  At leading order in $\tau$, we obtain
  \eqn\BoundaryLayer{\eqalign{
   {dA \over d\tau} &= {1 \over 4 {\root 6 \of {12}}}
     \left[ {-A^2 + D^2 \over BD} \right]  \cr
   {dB \over d\tau} &= {1 \over 4 {\root 6 \of {12}}}
     \left[ {A^2 + D^2 \over AD} - {C \over B} \right]  \cr
   {dC \over d\tau} &= {1 \over 4 {\root 6 \of {12}}}
     \left[ {C^2 \over B^2} \right] \cr
   {dD \over d\tau} &= {1 \over 2 {\root 6 \of {12}}}
     \left[ {A^2 - D^2 \over AB} \right] \,.
  }}
 The only regular solution is $A = D = (const)$ and $B = C = {1
\over 4 {\root 6 \of {12}}} \tau$.  We do not have a proof, but we
expect it is possible to match this ``boundary layer'' solution
smoothly onto the straightforward perturbation theory at $\tau \sim
C_\infty$, leading to a uniform approximation to a two-parameter
family of solutions.  One parameter is $C_\infty$, and we can get the
other by rigidly rescaling the whole metric.

To summarize, we have developed perturbation theory around the
solutions \coeffs\ and \finmet.  If we parametrize our solutions with
the radius $R_1$ of the circle at infinity and the radius $R_2$ of the
unshrunk ${\bf S}^3$, then we can describe slivers of the parameter
space with $R_1/R_2$ small or with $R_1/R_2 \approx 3/\sqrt{2}$.

\newsec{Effective Lagrangian approach}
In this section we rederive using a different method the first order
equations \BPSeqnsa\ that we obtained previously by imposing the existence of a
$G_2$-structure. In this other method we first find the equations of
motion which follow from imposing that the metric ansatz be Ricci flat.
If we express the metric ansatz \metri\siebenbein\ in terms of the new
functions
$\alpha(t)$, $\beta(t)$, $\gamma(t)$, and $\delta(t)$ as
\eqn\metric{\eqalign{
ds^2 = e^{4 \alpha + 4 \beta + 2\gamma + 2\delta} dt^2
&+ e^{2 \alpha} \big[ (\sigma_1-\Sigma_1)^2 + (\sigma_2-\Sigma_2)^2 \big]
+ e^{2 \delta} (\sigma_3-\Sigma_3)^2 + \cr
&+ e^{2 \beta} \big[ (\sigma_1+\Sigma_1)^2 + (\sigma_2+\Sigma_2)^2 \big]
+ e^{2 \gamma} (\sigma_3+\Sigma_3)^2 \,,}}
then a convenient way of obtaining the Ricci flatness equations is to
realize that these follow from the equations of motion which are
derived from the Einstein-Hilbert action  $S=\int dx^7 \sqrt{g}\;
R$. Therefore, by computing the Einstein-Hilbert action on the ansatz
one can reduce the problem to an effective quantum mechanics
problem. Evaluating the action on the ansatz \metric\ leads to the
following $0+1$-dimensional effective Lagrangian\foot{As usual in
this approach we have replaced the second order terms by first order
ones by
integrating by parts.}
\eqn\lagrangian{\eqalign{
L_{\hbox{eff}} = T - V &= 16 (\alpha')^2 + 64 \alpha' \beta'
+ 16 (\beta')^2 + 32 \alpha' \gamma'+ 32 \beta' \gamma'
+ 32 \alpha' \delta'+ 32 \beta' \delta' + 16 \gamma' \delta' \cr
& -2 e^{2(3 \alpha + \beta + \gamma)} -2 e^{2(\alpha + 3 \beta + \gamma)}
+4 e^{4 \alpha + 4\beta + 2\gamma}
+8 e^{2(2 \alpha + \beta + \gamma + \delta)} \cr
&+ 8 e^{2(\alpha + 2 \beta + \gamma + \delta)}
-2 e^{2(\alpha + \beta + \gamma + 2 \delta)}
- e^{4 \alpha + 4 \gamma + 2 \delta} - e^{4 \beta + 4 \gamma + 2 \delta}
}}
where a prime stands for a derivative with respect to the ``time''
coordinate $t$.
The first line in this expression should be understood
as the kinetic term for the scalar fields
$\alpha_i = (\alpha, \beta, \gamma, \delta)$. It can be written as
\eqn\kterm{T = {1 \over 2} G_{ij}
{\partial \alpha_i \over \partial t}
{\partial \alpha_j \over \partial t} }
with the following constant scalar-manifold metric
\eqn\scalarmet{
G_{ij}=\pmatrix{32&64&32&32\cr
           64&32&32&32\cr
           32&32&0&16\cr
           32&32&16&0}.}
The last two lines in the effective Lagrangian \lagrangian\
represent the scalar potential. One can derive first order equations
from \lagrangian\ if one manages to write the scalar potential of the
auxiliary Lagrangian \lagrangian\ in terms of a superpotential
\eqn\scalarV{ V = - {1 \over 2} G^{ij}
{\partial W \over \partial \alpha_i}
{\partial W \over \partial \alpha_j}.}
This can be accomplished and the corresponding superpotential reads
\eqn\superw{
W = 2 \sqrt{2} \big[
2 e^{3 \alpha + \beta + \gamma} + 2 e^{\alpha + 3 \beta + \gamma}
+ e^{2 \alpha + 2 \gamma + \delta}
- e^{2 \beta + 2 \gamma + \delta}
+ 2 e^{\alpha + \beta + \gamma + 2 \delta}
\big].}
Therefore, the following first order BPS-like equation
\eqn\BPS{ {\partial \alpha^i \over \partial t} =
G^{ij} {\partial W \over \partial \alpha^j} }
along with the constraint $T + V =0$ guarantees that any solution of
\BPS\ is a solution of the Ricci flatness equations.
By rewriting these first order equations in terms of the original variables
\eqn\newvars{
A = e^{\alpha}, \quad
B = e^{\beta}, \quad
C = e^{\gamma}, \quad
D = e^{\delta}, \quad
dr = \sqrt{2} e^{2 \alpha + 2 \beta + 2 \gamma + \delta} dt}
we obtain the first order equations \BPSeqns\ which we obtained before
by a more direct method. In the new variables $A$, $B$, $C$, and $D$,
the superpotential \superw\ takes the form of a homogeneous polynomial
of degree five:
\eqn\superww{
W = 2 \sqrt{2}
\big[ 2 A^3 BC + 2 AB^3C + A^2C^2D - B^2C^2D + 2 ABCD^2 \big].}
Every solution to the BPS-like equation \BPS\ should interpolate
between critical points of the superpotential. Since in our case $W$
is a homogeneous polynomial, the only critical points are $W=0$
and $W=\infty.$ Therefore, for any given solution the image of $W(r)$
should be given by a semi-infinite line going from $W=0$ to $W=\infty$.
The space of such solutions is parametrized by three real numbers
(not including the trivial shift of the radial variable $r$),
which can be identified, for example, with the value of $C$ at
$W=\infty$ and with values of $A$ and $B$ at $W=0$.
Roughly speaking, these parameters represent the size of
the ${\bf S}^1$ fiber and the volumes of two $3$-spheres, respectively.
We are interested in solutions which have $B=0$ in
the interior.\foot{It follows from the extremality conditions that
for non-zero $A$, the value of $C$ has to vanish at $W=0$ as well.}
The solutions satisfying this extra condition are parametrized
in general by two real parameters, in agreement with our analysis
in section $3$.
In the effective Lagrangian approach one has
to prove that the metric has reduced holonomy. Luckily, we have
already constructed the corresponding $G_2$-structure for this metric
and have in fact shown that the one-parameter solution
\coeffs\ has $G_2$ holonomy.
\newsec{Type IIA Reduction and Wrapped D6-Branes}
The $G_2$ holonomy metric is a solution of eleven-dimensional
supergravity. It can be used to describe a four-dimensional
vacuum\foot{Since the space is non-compact gravity lives in eleven dimensions.}
with four-dimensional ${\cal N}=1$ supersymmetry
of the Type ${\bf R}^{1,3}\times
{\bf X}_7$ where ${\bf X}_7$ is the seven manifold under
consideration. The metric we found \metrcol\coeffs\ has a $U(1)$
isometry which act by shifts on an angular coordinate. Therefore, we
can reduce the solution along this $U(1)$ isometry to obtain a
Type IIA solution by using
\eqn\reducea{
ds_{11}^2=e^{-2\phi/3}ds_{10}^2+e^{4\phi/3}(dx_{11}+C_\mu dx^\mu)^2,}
where $\phi$ and $C$ are respectively the Type IIA dilaton and
Ramond-Ramond one-form gauge field.
Thus, reducing the solution we found in \metrcol\coeffs\ and
identifying $x_{11}=\psi_2$ one obtains the following Type IIA solution
\eqn\typeIIsol{\eqalign{
ds_{10}^2 & = r_0^{1/2}C \left(dx^2_{1,3} + A^2 \left( (g^1)^2 +
(g^2)^2 \right) +
B^2 \left( (g^3)^2 + (g^4)^2 \right) + D^2 (g^5)^2 \right) +
r_0^{1/2} dr^2/C ~, \cr
e^\phi & = r_0^{3/4}(C)^{3 \over 2} ~,\cr
F_2 &= \left( \sin \theta_1 d\phi_1 \wedge d\theta_1 -
\sin \theta_2 d\phi_2 \wedge d\theta_2  \right) ~,}}
where the $g^i$'s are given by \basis .
This solution describes a D6-brane wrapping the ${\bf S}^3$ in the
deformed conifold geometry. At infinity the Type IIA metric becomes
that of the singular conifold and the flux is through the ${\bf S}^2$
surrounding the wrapped D6-brane. Moreover, the dilaton is constant
at infinity.
One can also analyze the solution in the interior.
For $r - 9r_0/2 = \epsilon \to 0$ the string coupling goes to zero
$e^\varphi \sim \epsilon^{3 \over 4}$ whereas the curvature blows up as
${\cal R} \sim \epsilon^{- {3 \over 2}}$ just like in the near horizon region
of a flat D6-brane. This means that
classical supergravity is valid for sufficiently
large radius. However, the singularity in the interior is the same as the
one of flat D6 branes, as expected. On the other hand the dilaton
continuously decreases from a finite value at infinity, set by the radius
$r_0$, to zero, so that for small $r_0$ classical string theory is valid
everywhere.
The global geometry is that of a warped product of flat Minkowski
space and a non-compact
space, $Y_6$, which for large radius is simply the conifold since the
backreaction
of the wrapped D6 brane becomes less and less important. In the interior
however the  backreaction induces changes of $Y_6$ away from the conifold
geometry.
At $r \to 9 r_0/2$ the  ${\bf S}^2$ has shrunk
to zero size whereas an ${\bf S}^3$ of finite size remains. Note, that this
behavior is similar to that of the deformed conifold but the two metrics
are different.
One can mod out the original eleven-dimensional metric  by
the following $Z_N$ action:
\eqn\singzn{Z_N \colon \psi_2 \to \psi_2+ {\pi \over N}}
The fixed points of this $Z_N$ action are located on the ${\bf S}^3$,
where the size of the circle parameterized by $\psi_2$ goes to zero.
Thus, the local geometry at $r \approx 9 r_0/2$ is singular,
with $A_{N-1}$ singularity fibered over ${\bf S}^3$,
the so-called singular quotient \AMV.
After compactification to Type IIA theory it describes $N$ coincident
D6-branes wrapped on the supersymmetric ${\bf S}^3$  of the deformed
conifold.
\newsec{A $U(1)$ Puzzle}
 Having established in section $5$ that the $G_2$ manifold given by
\metricres\coeffs\ reduces in the Type IIA language to a collection of
wrapped D6-branes
we analyze in this section the spectrum of massless fields both in
Type IIA and in M~theory and we point out that there is an apparent
discrepancy in the spectrum.
In the Type IIA side we expect to have a $U(N)$ gauge theory living on the
brane arising from the usual massless open strings of the N
D6-branes. The only subtlety
here is that the D6-brane normal bundle is non-trivial and the gauge
theory on the branes is topologically twisted. Since the
supersymmetric ${\bf S}^3$ is rigid the gauge theory one obtains at
low energies is a four-dimensional ${\cal N}=1$ $U(N)$ gauge
theory. The M~theory origin of the $SU(N)$ part of the gauge group is
familiar. The M~theory geometry is that of an $A_{N-1}$ singularity
fibered over ${\bf S}^3$ and the massless gauge fields come from
membranes wrapping the shrunken cycles of the $A_{N-1}$
singularity. So we now search for the  origin of the overall
$U(1)$ in M~theory.

In flat space, the $U(1)$ gauge field on a D6-brane is realized in
M~theory by performing a Kaluza-Klein reduction of the
eleven-dimensional supergravity three-form gauge field $C$
on the harmonic two-form $\eta$ of
the Taub-NUT geometry. This gives rise to a $U(1)$ gauge field since
we take  $C=A_\mu \wedge \eta$ and $A$ becomes a fluctuating dynamical
field since $\eta$ is a $L^2$-normalizable harmonic
two-form. Likewise, the $U(1)$ mode on  a D6-brane which wraps the deformed
conifold geometry should come from reducing $C$ along the
$L^2$-normalizable harmonic two-form $\eta$ of the geometry which describes the
wrapped D6-brane in M~theory and which is given by \metricres\coeffs .
We therefore must investigate whether our $G_2$ holonomy manifold admits an
$L^2$-normalizable harmonic two-form $\eta$. Our space
is homotopic to ${\bf R}^4\times{\bf S}^3$ which implies that our
two-form, if it exists, must be topologically trivial,
i.e. exact.
Moreover, this two-form must have the full symmetry of
the metric, namely $SU(2)\times SU(2)\times U(1)\times Z_2$. Such
a form can be constructed from the one-form which integrates to a
constant along the finite size ${\bf S}^1$ at infinity. Even though
this one-form cannot be extended as a closed one-form in the interior
it must be harmonic. Therefore, the
candidate two-form is given by
\eqn\candi{
\eta=d(F(r)(\sigma_3+\Sigma_3)).}
It is a straightforward exercise to show that the non-singular
harmonic two-form corresponds to $F=C^2$ and that in fact the one-form
$F(r)(\sigma_3+\Sigma_3)$ is itself harmonic.
It turns out that this one-form has an interesting geometrical
origin.
As mentioned earlier the metric \metri\siebenbein\ has
a $U(1)$ isometry whose Killing vector is given by
$\partial_\psi+\partial_{\widetilde{\psi}}$. It is a well known
fact that given a $U(1)$ Killing vector $K=K^{M}\partial_M$ on a
Ricci flat manifold, the dual one-form $K=K_M dx^M$ with
\eqn\kill{
K_M=g_{MN}K^{N},}
where $g_{MN}$ is the Ricci flat metric is harmonic. Therefore, the
candidate two-form is given by
\eqn\harmtwo{
\eta= d \big( C^2(\sigma_3+\Sigma_3) \big).}
In order for this form to give rise to a $U(1)$ gauge field in four
dimensions it has to be $L^2$-normalizable. Unfortunately,
 its norm is badly divergent
\eqn\norm{
||\eta||=\int_{{\bf X}_7}\eta\wedge *\eta\sim \Lambda^2,}
where $\Lambda$ is an IR regulator. This suggests that the $U(1)$ mode
obtained by Kaluza-Klein reduction is a parameter and not a
fluctuating dynamical field. It would be nice to reconcile this
calculation with the Type IIA expectation.

One can likewise compute the norm of the associative three-form. Reducing
the supergravity three-form $C$ along it gives rise to a scalar
mode. However, since $\Phi$ is not $L^2$-normalizable
\eqn\normass{
||\Phi||=\int_{{\bf X}_7}\Phi\wedge *\Phi\sim \Lambda^6,}
the scalar mode is a real parameter. This real parameter combines
with a real scalar parameter from the metric into a complex scalar
which is part of a four-dimensional ${\cal N}=1$ chiral multiplet.

It is possible to find other harmonic two-forms $\eta$ preserving the
$SU(2) \times SU(2) \times U(1) \times Z_2$ symmetry by projecting
onto irreps of $G_2$.  That is, we can demand $\eta \wedge \Phi =
2\eta$ or $\eta \wedge \Phi = -\eta$ to restrict to the ${\bf 7}$ or
${\bf 14}$ of $G_2$.  Then it turns out that $\eta$ is harmonic if it
is closed.  The result is that there is one more harmonic form in the
${\bf 7}$, but it is singular at $r=9r_0/2$; and there is one regular
harmonic form in the ${\bf 14}$, but it is highly non-normalizable at
infinity.  Neither of these two-forms seems to resolve the $U(1)$
puzzle.

\newsec{Summing up Membrane Instantons}
Another interesting aspect of the IR dynamics of the four-dimensional
effective $\CN=1$ gauge theory is the superpotential generated by instantons.
In this section we explain the origin of the effective superpotential
directly in M~theory on the $G_2$ manifold $X$.
Since there are no background fluxes
or branes in the M~theory compactification on $X$, the
effective superpotential $W$ in M~theory is generated only by
instantons corresponding to Euclidean membranes wrapped
on supersymmetric 3-cycles in $X$. Note, there are no
five-brane instantons since $G_2$-holonomy manifolds in general
do not have supersymmetric 6-cycles. In fact,
in our model\foot{
The 7-manifold $X$ of $G_2$ holonomy constructed in this paper
has $b_3=1$. Therefore, it has one modulus, which could be
interpreted as a scalar component of a chiral multiplet in
the four-dimensional $\CN=1$ effective field theory \sugrakk.
However, as we explained in the previous section,
this field is non-dynamical since the corresponding harmonic
form is not $L^2$-normalizable. In this sense, in our model
$W$ is a function of the coupling constant. However,
following the notations of \refs{\Vafa,\AKV}, here we
refer to the function $W$ as a superpotential,
bearing in mind applications to more general models.} we have $H_6(X)=0$.

The problem of counting the contributions of multiple covers
of membrane instantons usually prevents one from doing the calculation
beyond the one-instanton approximation \HM.
Here, summing up the entire instanton series for our model,
we demonstrate how geometric dualities conjectured by Vafa \Vafa\
open an avenue for such calculations, reducing the problem to counting
world-sheet instantons in type IIA string theory. This calculation provides
a further evidence for the membrane multiple cover formula proposed
by Ooguri and Vafa \OV, at least in the case of rational homology spheres:
\eqn\multicover{c_n = {1 \over n^2}}

Depending on the orientation of the supersymmetric 3-cycle
with respect to the $U(1)$ fiber (``M~theory circle'')
each membrane instanton can become in type IIA:

$i)$ an open string world-sheet instanton;

$ii)$ a D2-brane instanton;

$iii)$ a closed string world-sheet instanton.

The first option is realized when the 3-cycle can be represented
as a $U(1)$-bundle over a disk, so that the size of the fiber
goes to zero at the boundary of the disk. This possibility was
explained in a recent paper \AKV.
For the effects of open string instantons see \refs{\KKLM, \AganagicV, \AKV}.
The second option, when a membrane instanton reduces to a D2-brane
instanton is trivial, and occurs when the
$U(1)$ is ``orthogonal'' to the 3-cycle.
The last option occurs when the 3-cycle can be represented as
a non-trivial Hopf fibration of $U(1)$ over ${\bf S}^2$.
In this case, a membrane instanton is reduced to the genus zero
closed string world-sheet instanton.
Notice, that in all cases, a particular membrane instanton
is reduced to the corresponding instanton in type IIA theory.
This property will allow us to make an identification of their
contributions to the superpotential instanton-by-instanton.

The contribution of a single membrane instanton to the non-perturbative
superpotential was analyzed by Harvey and Moore \HM:
\eqn\wm{\Delta W \sim \vert H_1 (\Sigma,Z) \vert
\exp \Big( - \int_{\Sigma} (\Phi + iC) \Big) ~, }
where, $\Sigma$ denotes a supersymmetric 3-cycle.
Our space $X$ has only one compact supersymmetric 3-cycle $\Sigma \cong {\bf
S}^3$
calibrated with respect to the three-form \assocform, so the only
possible membrane instantons are those, which wrap $\Sigma \subset X$.
Following \AMV, we may consider a non-singular quotient of
our metric by the group $Z_N$ which acts freely on ${\bf S}^3$.
Hence, in this case $\Sigma = {\bf S}^3 / Z_N$ and  $H_1 (\Sigma,Z) = Z_N$.
If we denote the complexified
volume of $\Sigma$ by $z = \int_{\Sigma} (\Phi + i C)$, we can
write the sum over multiple covers of $\Sigma$ in the form:
\eqn\wc{W (z) = N \sum_{n=1}^{\infty} c_n \exp (- n z)}
where $c_n$ are numerical coefficients that can be obtained
using the dual type IIA descriptions.
One of the dual descriptions involves type IIA string theory
on the resolved conifold geometry $Y_6$ given by an
$\CO(-1) \oplus \CO(-1)$ bundle over $\P^1$ with $N$ units of
Ramond-Ramond two-form flux $F$ through the basic 2-cycle \refs{\Vafa,\AMV}.
In this picture
the effective ${\cal N}=1$ four-dimensional theory has a tree-level
superpotential generated by the Ramond-Ramond two-form flux \refs{\TV,\Gukov}:
\eqn\wkk{W = \int_{Y_6} F \wedge {\cal K} \wedge {\cal K}}
where the two-form ${\cal K}$ is obtained by integrating the associative
three-form $\Phi$ over the ${\bf S}^1$ fiber as outlined in \AMV.
The two-form ${\cal K}$ agrees with the K\"ahler form on the resolved
conifold before we turn on Ramond-Ramond flux.
This expression for the superpotential,
which is valid only in the large volume limit,
is corrected by world-sheet instantons.
The resulting superpotential can be written in
terms of genus zero closed topological string amplitude $F_0(t)$ \Vafa:
\eqn\wkkk{\eqalign{ W (t) &= \int_{Y_6}
F \wedge {\cal K} \wedge {\cal K}
+ {\rm instantons} = N {\partial F_0(t) \over \partial t} = \cr
&= N \sum_{n=1}^{\infty} {e^{-nt} \over n^2} } ~,}
where $t = \int_{P^1} {\cal K}$ denotes the size of the basic
two-cycle.

The same expression for the superpotential (modulo an ambiguous
polynomial piece \refs{\Vafa,\lastAV}) can be obtained from the disk
instanton calculation in another type IIA dual description including
D6-branes \refs{\KKLM,\AganagicV,\AKV}.
The parameters in the Type IIA compactification
on $Y_6$ and in the M~theory compactification on $X$,
in the large volume limit ({\it i.e.} when $z$ is large) should be
identified as $z \approx t$ \AMV. If we now compare the contribution
of every single instanton in \wkkk\ and in \wc,
we find that the two expressions agree, provided the coefficients
$c_n$ are given by the formula \multicover\
for multiple membrane wrapping in M~theory, as proposed in \OV.
Hence, when $\Sigma$ is a rational homology 3-sphere it is natural
to conjecture the following general formula for the effective
superpotential that includes the contributions of multiple covers:
\eqn\wrational{W = \sum {\vert H_1 (\Sigma,Z) \vert \over n^2}
\exp \Big( - n \int_{\Sigma} (\Phi + iC) \Big)}
This should be compared with the coefficients $c_n = n^{-3}$ in the genus
zero topological string partition function $F_0 (t)$
that account for multiple covers by fundamental strings \refs{\cand,\aspmor}.
It is amusing to check formula \wrational\ in more general
compactifications of M~theory on $G_2$ manifolds. For example,
a similar analysis\foot{We thank C.~Vafa for extensive
discussions on this point.} for quotients by dihedral groups suggests
that one has to sum over all membrane topologies in \wrational\
in order to reproduce the factor $(N \pm 4)$ in the corresponding
Type IIA calculation \wkkk. In fact, the 4 in this expression
comes as before from the formula \wm\
since $H_1 (\Sigma,\Z)$ is an abelian group of
order 4 for $\Sigma \cong {\bf S}^3 / {\bf D}_N$.
The leading contribution $N$ comes from bound states of two
basic membrane instantons, which transform into each other
under the action of one of the generators of the dihedral
group ${\bf D}_N$. These membrane instantons
become genus zero world-sheet instantons after reduction to Type IIA
string theory \SV. It would be very interesting to extend this analysis
to more general models.
\newsec{A More General Ansatz}
In view of the numerous applications $G_2$ holonomy metrics
and the scarcity of explicitly known metrics
it is important to search for more examples.
Our ansatz \metri\siebenbein\ allows for a straighforward
generalization by introducing
six independent functions which breaks the symmetry from
$SU(2) \times SU(2) \times U(1) \times Z_2$
down to $SU(2) \times SU(2) \times Z_2$.
The metric is given by
\eqn\metricgen{
ds^2 = \sum_{a=1}^7 e^a\otimes e^a,}
in terms of the following vielbeins
\eqn\siebenbeinred{\eqalign{
e^1 & = A_1(r) (\sigma_1-\Sigma_1) ~,~~e^2 = A_2(r) (\sigma_2-\Sigma_2) ~, \cr
e^3 & = A_3(r) (\sigma_3-\Sigma_3) ~,~~e^4 = B_1(r) (\sigma_1+\Sigma_1) ~, \cr
e^5 & = B_2(r) (\sigma_2+\Sigma_2) ~,~~e^6 = B_3(r) (\sigma_3+\Sigma_3) ~, \cr
e^7 & = dr.}}
One can construct the associative three-form $\Phi$ as in section
$2$. Imposing closure and co-closure of the $G_2$-structure yields the
following system of first order differential equations\foot{After
completing this work, we were informed about \toappear, where these
equations and some aspects of their solutions have been discussed.
We would like to thank M. Cveti\v{c}, G.W. Gibbons, H. L\"u and C.N. Pope
for sharing their results prior to publication.}
\eqn\genBPS{\eqalign{
{d A_1 \over dr} & = - {1 \over 4} \left[ {A_1^2-A_3^2-B_2^2 \over A_3 B_2} +
   {A_1^2 - A_2^2 - B_3^2 \over A_2 B_3} \right] \cr
{d A_2 \over dr} & = - {1 \over 4} \left[ {A_2^2-A_3^2-B_1^2 \over A_3 B_1} +
   {A_2^2 - A_1^2 - B_3^2 \over A_1 B_3} \right] \cr
{d A_3 \over dr} & = - {1 \over 4} \left[ {A_3^2-A_2^2-B_1^2 \over A_2 B_1} +
   {A_3^2 - A_1^2 - B_2^2 \over A_1 B_2} \right] \cr
{d B_1 \over dr} & = {1 \over 4} \left[ {A_2^2+A_3^2-B_1^2 \over A_2 A_3} +
   {B_1^2 - B_2^2 - B_3^2 \over B_2 B_3} \right] \cr
{d B_2 \over dr} & = {1 \over 4} \left[ {A_1^2+A_3^2-B_2^2 \over A_1 A_3} +
   {B_2^2 - B_1^2 - B_3^2 \over B_1 B_3} \right] \cr
{d B_3 \over dr} & = {1 \over 4} \left[ {A_1^2+A_2^2-B_3^2 \over A_1 A_2} +
   {B_3^2 - B_1^2 - B_2^2 \over B_1 B_2} \right] }}
We note that these equations can also be found by the effective
Lagrangian method we used in section $4$.

Unfortunately, we were not able to find closed solutions to this
complicated set of differential equations. It is worthwhile however
to end this section with a speculation on the solutions of \genBPS.
At several places we mentioned the similarity of our solution with the
Taub-NUT metric which described the M~theory lift of D6 branes.
Taub-NUT itself has an $SU(2) \times U(1)$ symmetry, but there exists
another four-dimensional self-dual metric without the $U(1)$ symmetry
which has an interesting interpretation in M~theory. It is the
Atiyah-Hitchin metric
\nref\atiyah{M.~F.~Atiyah and N.~J.~Hitchin,
``Low-Energy Scattering Of Nonabelian Monopoles,''
Phys.\ Lett.\ A {\bf 107}, 21 (1985).}
\atiyah\
which describes the uplift of an ${\cal O}6^{-}$
orientifold plane in Type IIA
\nref\seiwit{N.~Seiberg and E.~Witten,
``Gauge dynamics and compactification to three dimensions,''
hep-th/9607163.}%
\nref\asen{A.~Sen,
``A note on enhanced gauge symmetries in M- and string theory,''
JHEP {\bf 9709}, 001 (1997).}%
\refs{\seiwit,\asen}.
Hence, it is tempting to conjecture that
the solutions of \genBPS, for certain values of the parameters, describe an
${\cal O}6^{-}$ plane --- possibly in the presence of additional D6 branes
--- wrapped on ${\bf S^3}$ of the deformed conifold.
Such backgrounds would be the
supergravity duals of ${\cal N} = 1$ supersymmetric Yang-Mills with
gauge group $SO(2 N)$.

\newsec{Discussion}
In this paper we found explicitly a new one-parameter family of
Ricci flat metrics,
which have
$G_2$ holonomy group and also the structure of
$U(1)$ bundle over the conifold geometry
$T^* {\bf S}^3$.
The topology of these metrics is that of the
spin bundle over ${\bf S}^3$. These metrics, unlike the previously known
asymptotically conical $G_2$ holonomy metrics, have a circle of
finite radius at infinity. The size of the circle has an
interpretation in string theory as the size of the Type IIA string
coupling constant.
This new metric, if used as a vacuum solution of M~theory, has a nice
interpretation as a Type IIA string theory solution. Once the $G_2$
holonomy metric is reduced to Type IIA along it provides the
supergravity description of a collection of D6-branes wrapping the
supersymmetric ${\bf S}^3$ in the deformed conifold geometry. As such
it plays a r\^ole as a supergravity dual to supersymmetric Yang-Mills
theory.

The metric we have constructed
pertains to the duality conjectured by Vafa \Vafa\
and recently geometrized in \refs{\AcharyaMTH,\AMV,\atwit}. This duality
states that Type IIA string theory with wrapped D6-branes on the
deformed conifold is dual to Type IIA string theory on the resolved
conifold with flux. In fact there are two sides of the duality with
flux related by the familiar flop transition. Since these Type IIA
backgrounds involve only the metric, dilaton and Ramond-Ramond one-form
gauge field
they have a purely geometric description in M~theory. Supersymmetry
dictates that the M~theory geometry is that of a manifold with $G_2$
holonomy. As shown in \refs{\AMV,\atwit} these dual theories are indeed
connected in M~theory since the three $G_2$ holonomy metrics
reside on
the same moduli space. Moreover, for arbitrary value of the
$\theta$-angle of the supergravity three-form one can in M~theory smoothly
interpolate between the three geometries without a phase transition.
According to \refs{\AKV,\lastAV}, each of these three phases
has a further discrete choice of behavior of the $G_2$ metric
at large distances related to the so-called framing
ambiguity.\foot{We thank C.~Vafa for pointing this out.}
It implies the existence of infinitely many $G_2$ metrics of the same
topology, which have the same behavior in the interior, but differ in
the choice of the $U(1)$ fiber at large distance. In this context, the new
metric constructed here should correspond to the canonical framing,
$p=0$ in the notations of \lastAV.

In this paper we have found one of these metrics, namely,
the one describing
 the deformed conifold geometry with branes.
Needles to say, it would
be very interesting to find the $G_2$ holonomy
metrics which upon
reduction to Type IIA describe the resolved conifold
with
flux and/or corresponding to the different choice of framing.
Finding these and analogous metrics will hopefully teach us --
among other things -- about compactifications on $G_2$ holonomy
manifolds and improve our understanding of
the type of dualities suggested by Vafa
\Vafa\ between branes and fluxes.

\centerline{\bf Acknowledgments}
We are grateful to Michael Atiyah,
Vadim Borokhov, Mirjam Cvetic, Igor Klebanov,
Chris Pope, and especially Cumrun Vafa and
Edward Witten for useful discussions.
The work of A.~Brandhuber and S.~Gubser is supported in
part by the DOE under grant No. DE-FG03-92ER40701.  The work of
J.~Gomis is supported in part by the National Science Foundation under
grant No. PHY99-07949 and by the DOE under grant
No. DE-FG03-92ER40701.  The research of S.~Gukov is supported in part
by the Caltech Discovery Fund, NSF grant No. PHY99-07949, grant RFBR
No. 01-01-00549, and the Russian President's grant No. 00-15-99296.

\vfill\eject
\appendix{I}{The Vielbein Agebra}
In this appendix we present the basic vielbein algebra required to
perform the computations in this paper. The metric ansatz we want to
consider is
\eqn\appe{
ds^2 = \sum_{a=1}^7 e^a\otimes e^a,}
with the following vielbeins
\eqn\appenvir{\eqalign{
e^1 & = A(r) (\sigma_1-\Sigma_1) ~,~~e^2 = A(r) (\sigma_2-\Sigma_2) ~, \cr
e^3 & = D(r) (\sigma_3-\Sigma_3) ~,~~e^4 = B(r) (\sigma_1+\Sigma_1) ~, \cr
e^5 & = B(r) (\sigma_2+\Sigma_2) ~,~~e^6 = C(r) (\sigma_3+\Sigma_3) ~, \cr
e^7 & = dr/C(r).}}
The exterior calculus of these vielbeins is
\eqn\alge{\eqalign{
de^1&=C{A^{\prime} \over A} e^7\wedge e^1+{A\over 2}\left({e^3\wedge e^5\over
BD}+{e^6\wedge e^2\over
AC}\right)\cr
de^2&=C{A^{\prime} \over A} e^7\wedge e^2+{A\over 2}\left({e^4\wedge e^3\over
BD}+{e^1\wedge e^6\over
AC}\right)\cr
de^3&=C{D^{\prime} \over D} e^7\wedge e^3+{D\over
2AB}\left({e^2\wedge e^4}+
{e^5\wedge e^1}\right)\cr
de^4&=C{B^{\prime} \over B} e^7\wedge e^4-{B\over
2}\left({e^2\wedge e^3\over AD}+
{e^5\wedge e^6\over BC}\right)\cr
de^5&=C{B^{\prime} \over B} e^7\wedge e^5-{B\over
2}\left({e^3\wedge e^1\over AD}+
{e^6\wedge e^4\over BC}\right)\cr
de^6&={C^{\prime}} e^7\wedge e^6-{C\over
2}\left({e^1\wedge e^2\over A^2}+
{e^4\wedge e^5\over B^2}\right)\cr
de^7&=0
},}
where $F^{\prime}\equiv {dF(r)\over dr}$.
\listrefs
\end